\documentclass[12pt]{article}

\usepackage{amsmath}
\usepackage{amssymb}
\usepackage{epsfig}
\usepackage{verbatim}  % to use the comment environment

\newcommand{\cD}{\mathcal{D}}

\newcommand{\cM}{\mathcal{M}}

\newcommand{\Mink}{\hbox{\openface M}}

\newtheorem{claim}{Claim}
\newtheorem{definition}{Definition}

\newcommand{\Om}{\Omega}

\newcommand{\ap}{\rightharpoonup}
\newcommand{\hP}{\hat{P}}

\newcommand{\ie}{\textit{i.e. }}

\font\openface=msbm10 at10pt

\def\Mink          {{\hbox{\openface M}}}

\def\Reals         {{\hbox{\openface R}}}

\begin{document}
\title {Quantum Histories and Quantum Gravity}

\author{Joe
Henson\footnote{Perimeter Institute, 31 Caroline St. N. , Waterloo, Canada N2L 2Y5.}}
\maketitle

\begin{abstract}
This paper reviews the histories approach to quantum mechanics.  This discussion is then applied to theories of quantum gravity.  It is argued that some of the quantum histories must approximate (in a suitable sense) to classical histories, if the correct classical regime is to be recovered. This observation has significance for the formulation of new theories (such as quantum gravity theories) as it puts a constraint on the kinematics, if the quantum/classical correspondence principle is to be preserved. Consequences for quantum gravity, particularly for Lorentz symmetry and the idea of ``emergent geometry'', are discussed.
\end{abstract}

 \vskip 1cm
%=====================================================

\section{Introduction}

Quantum histories are those entities summed over in a quantum path integral or ``state sum''.  How much of the physical interpretation that we give to classical histories (solutions of the classical equations of motion for the theory in question) is preserved in the quantum case?  Is it, at least, fair to say that some of the quantum histories must approximate, in some sense, to classical ones?  The answer has important consequences.  If this is true, then given a classical theory we could put some constraints on the space of quantum histories that could underlie it, even before we had written down the rest of the dynamics.

This may seem like a weak condition considering simple cases such as a single particle.  However, it becomes non-trivial if we consider emergence: the idea that the apparent properties that we (approximately) test for at the classical level are only effective descriptions of some more fundamental properties of the quantum system.  In that case there may be many possibilities for the quantum history space; it becomes more important to find methods for ruling out unsuitable ones.  This is relevant to quantum gravity (QG).  The idea that the spacetime manifold is merely an approximation to some deeper (and perhaps discrete) structure is a prevalent one in current QG research.  In that case, it is argued below, these matters have an impact on questions such as the preservation of Lorentz symmetry, and strategies for producing theories of emergent geometry in general.

However, in view of the current literature -- particularly in quantum gravity  -- the situation is unclear.  Certainly, quantum histories are quite unlike the classical solutions in some respects.  Many pieces of research are influenced by the fact that effective, macroscopic properties may be only cryptically related to the microscopic properties of a system.  As a result, it is sometimes argued that \textit{only} properties of quantum \textit{sums} of histories, rather than those of any single history, can be safely related to semiclassical properties (see \textit{e.g.} \cite{Markopoulou:2007ha,Markopoulou:2002ja}).  Also, in a certain view of quantum mechanics, the path integral exists only to calculate the evolution of the wave-function, and so the physical interpretation of the path integral is subsidiary to that of the wave-function.  Taking on this view encourages one to downplay questions about the properties of quantum histories.

Some of these statements seem to run contrary to other viewpoints on quantum mechanics.  In what might be called the ``histories viewpoint'', the results of measurements are directly related to properties of the histories \cite{Feynman:1948ur} \cite{Caves:1986is}(see also \cite{Sinha:1991cj} and references therein).  This allows an easy treatment of measurements distributed in time, and other insights into new and standard theories that would be difficult by other means.

Some closer analysis is necessary to reconcile, or choose between, the two views.  In the first part of this paper, some claims are made that explain how properties of quantum histories relate to measurements and to classical approximations.  The central claim argued for is that some of the quantum histories must approximate the classical histories in some reasonable class, if the correct classical regime is to be recovered.  A more precise idea of the required approximation will be given.  The relationship between symmetries and this approximation is discussed (section \ref{s:sym}), and it is explained how all this is consistent with the idea of effective descriptions (section \ref{s:effective}).  This leads to some physical conditions on effective descriptions: roughly, that the fundamental-effective correspondence should respect the symmetries of the classical theory in question.  In part \ref{p:qg}, this principle is applied to quantum gravity and theories of emergent spacetime, resulting in some constraints on such theories (coming from the requirement of local Lorentz invariance in the effective continuum description).

\begin{comment}
  In some contexts, arguments downplaying the properties of histories are justified.  Even the behaviour of a \textit{classical} statistical mechanical system can, on macroscopic scales, have only weak dependence on the microscopic dynamics.  In that sense, one might say that the individual configurations are of little importance. But in this classical case it is obvious that all of our measurement outcomes, at bottom, correspond to properties of the histories (in that case the ``histories'' are solutions of some classical equations of motion, or coarse-grainings of them).  In a similar way it is argued below that insights from an ``emergence'' or ``effective descriptions'' viewpoint do not run contrary to the main ideas of the histories viewpoint, even in the quantum case.  Effective descriptions of this type can be contained in a histories framework in a straightforward way.
\end{comment}

Some of the main points raised are not at all novel.  The general
viewpoint on quantum mechanics is present in Feynman's seminal paper \cite{Feynman:1948ur} and makes an appearance in other well-known discussions of quantum mechanics, e.g. the work of Bell \cite{Bell:1987}.  Hopefully, therefore, the conclusions of part \ref{p:hist} will be unsurprising to many researchers.  However, the application to quantum gravity and emergent dynamics does call on some details of this discussion.

At the outset, it should be noted that these arguments are in no way
reliant on a particular view of the measurement problem.  Neither is
it necessary to accept that a histories-based view of quantum
mechanics, as opposed to the state-vector based one, is somehow
primary.  It is only necessary that the described histories view of
a quantum mechanical theory is consistent; it need not be preferable
on some other grounds.  Consequently, the conclusions should not
only be relevant to those who are committed to a histories viewpoint
-- it is the possibility of such a viewpoint that is argued to have
important consequences.
\part{The Histories Approach}
\label{p:hist}

The purpose of this first part of the paper is to review some basic facts about the histories approach to quantum mechanics, and to justify some claims about histories and measurements in semi-classical situations.  When a specific example is called for, the quantum mechanical theory of a single non-relativistic particle will be used.  Although this example is useful for explanation, the points made will mostly be relevant to other systems as well.  Where it is not obvious, any important differences will be stated in discussion.

\section{Histories and measurements}
\label{s:hist}

The predictions of quantum mechanics come in the form of conditional probabilities of some experimental outcomes given some preparatory observations.  All of these can be computed from a joint probability distribution over the outcomes of all of the measurements.  Below, unless otherwise noted, we will use the Heisenberg picture, and measurements will be projective Von Neumann measurements.  We will consider making measurements at $N$ separate times $t_1,t_2,...,t_i,...,t_N$ on a single non-relativistic particle.  At time $t_i$ we test whether the position of the particle has some value in the set $\Delta_i$.  Let us define the corresponding (classical) observable $P_i\bigl(x(t_i)\bigr) := 1$ when $x(t_i) \in \Delta_i$ and $0$ otherwise.  In the operator formalism, the joint probability for passing all of the tests takes the following form:
\begin{equation}
\label{e::operator}
\Pr(x(t_i) \in \Delta_i\, \forall \, i \bigr) = \parallel
U(t_0,T) \, \hP_n \,  ... \, \hP_2 \, \hP_1\, |\psi_0 \rangle \parallel^2,
\end{equation}
where $|\psi_0 \rangle$ is the initial wave vector, $U(t',t)$ is the unitary evolution operator between times $t$ and $t'$, and from our definition of $P_i$,
\begin{equation}
\hP_i = U^\dagger(t_0,t_i) \:
\int_{\Delta_i} dx  |x \rangle \langle x|
\: U(t_0,t_i).
\end{equation}
The time evolution to the final time $T$ is included for later convenience. Using a path integral formulation, it is possible to write down another form for the joint probability.  First let us give the probability amplitude for the particle to pass all of these tests and end up at some position $x_T$ at a final time $T \geq t_n$:
\begin{equation}
\label{e::amp}
\Phi\bigl(x(t_i) \in \Delta_i\, \forall \, i ; x_T,T \bigr)
=\int_{t_0}^{x_T,T} \cD x(t) \, \Bigl(\prod_{i=1}^{N} P_i\bigl(x(t_i)\bigr) \Bigr)
e^{(i/\hslash) S[x(t)]} \psi_0(x(t_0)) \; ,
\end{equation}
where the limits of the path integral mean that we sum over all paths from an initial time $t_0$ to final time $T$ with a specific final position $x_T$, $\psi_0(x):=\langle x|\psi_0 \rangle$ is the initial wave function, and $S[x(t)]$ is the action of the path or ``history'' $x(t)$.   The joint probability to pass all the tests is
\begin{equation}
\label{e::probdist}
\Pr(x(t_i) \in \Delta_i\, \forall \, i)
=\int dx_T \, \Bigl \arrowvert \Phi\bigl(x(t_i) \in \Delta_i\, \forall \, i ; x_T,T \bigr)
\Bigr \arrowvert ^2 \; .
\end{equation}
This\footnote{The ``incoherent sum'' (\ie the sum outside the square) over final positions in eqn.(\ref{e::probdist}) is necessary for agreement with eqn.(\ref{e::operator}) , and follows from the use of the inner product in (\ref{e::operator}) -- the value of the final time $T$ has no effect on the result as is easily seen from   (\ref{e::operator}).} removes any direct reference to Hilbert space, operators and state vectors from our joint probability expression.  Similarly we can find the probability for any combination of passed and failed tests, and from this we can make a probability distribution on the space of all possible sequences of outcomes.  For measurements in the position (generalised) basis, and for dynamics that is local and Markovian in the appropriate sense \cite{Zinn-Justin:2005}, this is equivalent to the state-vector expression.  These results follow from repeated application of Feynman's prescription for calculating the probability of an outcome of one measurement with path integrals \cite{Feynman:1965}, as explained in \cite{Caves:1986is}
\footnote{This is a special case of Caves' prescription, which allowed amplitudes for detection as a function of position (in the operator formalism, POVM measurements rather than the projection measurements used here for simplicity).}.

Now let us consider the space of histories summed over in eqn.(\ref{e::probdist}), which will be called the \textit{history space} $\Omega$.  Each set of experimental outcomes is codified as a proposition about the system (a particle's position in this case), such as ``$x(t_i) \in \Delta_i\, \forall \, i$''.  This proposition corresponds to a possible property of a history $x(t)$ in $\Omega$. Also, it corresponds to a subset of the history space $\Omega$, namely, the set of histories that have that property.  For instance, let $\Gamma_i=\{x(t) \in \Omega \,:\, x(t_i) \in \Delta_i\}$.  Then $\Gamma= \bigcap_{i=1}^N \Gamma_i=\{x(t) \in \Omega \,:\, x(t_i) \in \Delta_i\, \forall \, i\}$ is the set that corresponds to passing all the tests.  This is the correspondence between measurement outcomes and properties of histories that we will be interested in throughout the following.

The joint probability can be rewritten using this idea.  Note that  $\prod_{i=1}^N P_i\bigl(x(t_i)\bigr)=1$ for any history $x(t) \in \Gamma$ and $0$ for other histories.  We now write the probability as a function of a subset of $\Omega$ rather than of the corresponding proposition.  From equations (\ref{e::amp}) and (\ref{e::probdist}),
\begin{equation}
\label{e::ampset}
\Phi(\Gamma ; x_T,T)
=\int_{t_0;\; x(t) \in \Gamma} ^{x_T,T} \cD x(t) \,
e^{(i/\hslash) S[x(t)]} \psi_0(x(t_0)) \; ,
 \end{equation}
where $x(t) \in \Gamma$ in the limits of the integral means that only paths in $\Gamma$ are summed over,  and the joint probability (which has been renamed) is
\begin{equation}
\label{e::probdistset}
\mu(\Gamma)
=\int dx_T \,  \Bigl \arrowvert \Phi(\Gamma ; x_T,T) \Bigr \arrowvert ^2 \; .
\end{equation}

\begin{figure*} \centering
\resizebox{5.2in}{1.8in}{\includegraphics{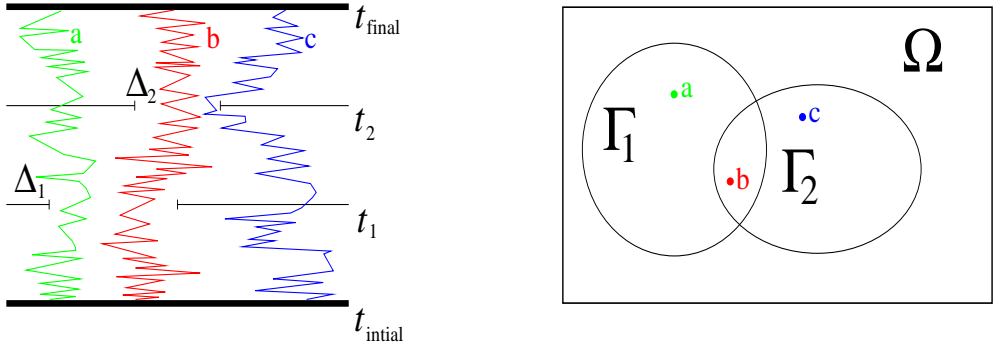}}
\caption{\small{The left hand diagram is a space-time diagram of some paths in a typical single particle path integral, and the right hand diagram shows the space of histories as a set diagram.  On the left time runs between initial and final times, and intervals $\Delta_1$ (at time $t_1$) and $\Delta_2$ (at time $t_2$) correspond to two measurements as described in the text.  They also correspond to sets of histories $\Gamma_1$ and $\Gamma_2$, as shown on the right.  Three histories are shown as examples on both sides.}\label{f:histories}} \end{figure*}

This form best illustrates the connection between experimental outcomes and properties of histories, shown in figure \ref{f:histories}.  Having made this connection, we can say the following: $\mu(\Gamma)$ is the probability that the history lies in the set $\Gamma$ -- assuming that we make this particular set of measurements \cite{Feynman:1948ur}.  This defines the probability distribution over the set of alternative outcomes for our chosen measurements. However, while we can uniquely define $\mu(\Gamma)$ for a set of histories $\Gamma$, $\mu$ need not satisfy the conditions to be a probability measure on $\Om$ \cite{Sorkin:1994dt}.  The double slit experiment provides a simple counter-example: conditioning on the final position of the particle being in a dark line on the screen, the probability of finding the particle to have passed through some slit (had we measured for that) is not the sum of the probabilities for passing through the left slit and passing through the right slit (had we measured for that), violating the probability sum rule.  That is why notation has been switched from $\Pr$ to $\mu$, which is sometimes called a ``quantum measure'' (as opposed to a probability measure) on $\Omega$ \cite{Sorkin:1994dt}, or the ``diagonal of the decoherence functional'' \cite{Hartle:1992as}.

One advantage of this path integral way of writing the probabilities is that it can easily be extended to measurements distributed in time, unlike the standard Hilbert space/operator formalism \cite{Caves:1986is}.  There is nothing to stop the definition of (and, in principle, measurement for) more general properties such as ``The particle was within the interval $\Delta$ at some time between times $t_1$ and $t_2$''.

Similar prescriptions can be made for other systems, given a suitable choice of a basis to function as the position basis does above, which will be called the \textit{histories basis} (it is important to note that not all bases are suitable in this regard).  Such a sum-over-histories formalism can be defined with greater or lesser levels of rigour for all of the standard quantum theories, and when the history space is discrete, the path integral (or state sum) is well-defined.  To summarise, we can make the following uncontroversial assertion:

\begin{claim} \label{c:outcomes}
The outcomes of the measurements in the histories basis
can be identified with properties of the quantum histories, \textit{i.e.} with subsets of the history space.
 \end{claim}

\subsection{Comparing quantum and classical histories}
\label{s:compare}

Claim \ref{c:outcomes} can be used to defend the central claim of this section:  that some quantum histories in our theory must approximate to classical histories in the corresponding classical theory.

Claim \ref{c:outcomes} does not assert that \textit{all} measurements correspond to properties of histories.  For instance, a measurement of the momentum $\hat{p}$ of our non-relativistic particle cannot always be derived from sequential measurements of its position, even on a coarse-grained level.  But it is possible to claim this in certain situations, namely, those in the domain of validity of the corresponding classical theory.  Otherwise, we would have no reason to give $\hat{p}$ the same name as the familiar classical momentum.  Deferring a fuller explanation until the next section, the claim is the following (note again that measurements here are restricted to projective Von-Neumann measurements):

\begin{claim} \label{c:basis}
 In a classical situation, all measurements can be represented by projection operators in one basis\footnote{The argument that projection operators (as opposed to POVMs) are general enough to describe classical measurements is made in \cite{Kofler:2007}.}.
 \end{claim}

From claims \ref{c:outcomes} and \ref{c:basis}, the next claim follows easily.

\begin{claim} \label{c:main}
 In a classical situation, the outcomes of all measurements
can be identified with properties of the histories (when a suitable basis is used as histories basis), \textit{i.e.} with subsets of the history space.
 \end{claim}

With the help of the correspondence principle, that quantum mechanics must agree with the predictions of classical mechanics in its domain of validity, this leads to our central claim.  Firstly it is useful to restate the principle as follows:

\paragraph{Correspondence principle:} \textit{If a quantum theory is to be consistent with its classical counterpart, then, for every classical history $\gamma^{\text{cl}}$ in some reasonable set, there must exist a classical situation in that quantum theory containing enough measurements to approximately reconstruct $\gamma^{\text{cl}}$.}

%could add: to approximately reconstruct
%"some portion" of $\gamma^{\text{cl}}$.
%But this is hardly an important distinction and seems to have no
%immediate consequences for the flow of the argument.
\paragraph{}Combining this principle with claim \ref{c:main} leads to the following:

\begin{claim} \label{c:approx}
For every classical history (in some physically reasonable subset of the classical histories), there must exist some quantum history that approximates it\footnote{It should be possible to strengthen this claim, and to demand not one approximating quantum history but some larger set.  Once the dynamics is specified, a set of histories that approximate the classical history should have a large quantum measure, and it seems unreasonable to expect a set composed of one history to do so.  However this weak form of the claim seems sufficient for the following applications.}.
\end{claim}

This claim will be a focus of the following discussion.  It will rule out some history spaces as good starting points for the quantisation of a given theory.  In standard cases our quantisation procedures ensure that this requirement is met, but this may no longer be true in a more general setting, as we will see.  Later we will use such a strategy to give insights into quantum gravity.

It remains to define and discuss the term ``classical situation'', and the notion of ``approximation'' referred to in the statement of the correspondence principle and in claim \ref{c:approx}.  This will justify claim \ref{c:basis}, from which the others follow.

How do we decide on the level of approximation necessary to maintain the correspondence principle?  For later convenience we will introduce the following notation: $\gamma \sim \gamma'$ if the two histories $\gamma$ and $\gamma'$ approximate to each other in the above sense.  The most basic requirement comes from observation.  Two histories cannot approximate to each other if they can be distinguished by an experiment which confirms the classical theory (with high probability).  Since it refers only to measurements in the domain of validity of the classical theory, this requirement can be useful \textit{prior} to the formulation of the quantum theory.  In practice we might demand slightly more:  that there will be similar classical situations possible in the theory for other possible classical solutions (although one must be careful not too demand too much \cite{Ashtekar:2005dm}).  This point is returned to in section \ref{s:sym}, where it is argued that a condition must be put on $\sim$ if we are to preserve the symmetries of the classical theory.

First, a common objection to comparing quantum and classical histories is noted.  Quantum histories can be shown to have several counter-intuitive and non-classical features.  A ``typical'' path in a position path integral\footnote{This idea of typicality is an informal notion.  It means a path which is a member of some reasonable high amplitude set, with ``reasonable'' having the same meaning as in the context of reasonable measurements \cite{Peres:1995}.} is nowhere differentiable, and has a fractal dimension.  In the continuum, the length along the path is typically infinite (these points are briefly reviewed in a quantum gravity context in \cite{Ambjorn:1998pz}).  In these ways the typical quantum histories are quite unlike classical histories.

\begin{figure*} \centering
\resizebox{2.2in}{2.0in}{\includegraphics{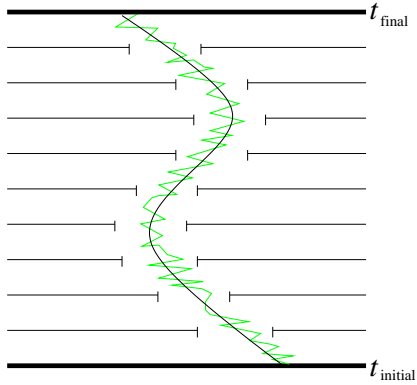}}
\caption{\small{A sketch informally illustrating a classical situation.  The sketch is a space-time diagram like the left hand side of figure \ref{f:histories}.  The black curve represents a classical history.  The light-coloured path represents a typical path in the path integral.  Measurements, corresponding to the spatial intervals represented by capped lines, are consistent with classical properties with high probability.  Such measurements must be of finite accuracy, and be separated by finite intervals in time.  However, with enough measurements the properties of the classical history are sufficiently well approximated.}\label{f:classical}} \end{figure*}

This gives good reason to be skeptical of applying classical thinking to all aspects of the path integral.  However, this argument does not speak against the claims made above.  It is impossible to measure the smoothness of the spacetime trajectory of a particle, or its total length; to do so one would have to measure with total precision the position of the particle at all times. Such measurements are quite unlike the measurements we perform to verify our classical theories in their domains of validity.  Claim \ref{c:approx} only states that there must be some quantum histories that share all the (coarse-grained) properties of a classical history that we could observe in a classical situation.  Paths may be unsmooth and of infinite length, but all the appropriate coarse-grained properties can be classical, as sketched in figure \ref{f:classical}.  This is the basic meaning of ``approximation'' in the claim; a lot hangs on the subtleties of this approximation, which will be examined later.

Claim \ref{c:approx} puts a condition on the properties of individual histories in the sum-over-histories, not on ``quantum sums'' of histories.  How are the properties of sets of histories different from those of the quantum histories themselves?  At the classical level, these collective properties are nothing more than \textit{some subset} of the properties of some quantum history: those properties common to all histories in the sum.  According to the claims, there are no properties at the level of the classical approximation \textit{additional} to those of the histories; the classical properties are not ``strongly emergent'' in that sense.

\subsection{Classical situations}
\label{s:classmeas}

A \textit{classical situation} is here defined as one in which no deviation from the classical theory is observed.  In QM (using the Heisenberg picture), we have a state, a dynamics and a set of projection measurements $M$ making up an ``experimental situation'' $\{|\Psi\rangle,U(t,t'),M\}$.

\begin{definition} \label{st:m}
A situation $\{|\Psi\rangle,U(t,t'),M\}$ is classical if, with high probability, none of the measurements in $M$ contradict the predictions of some solution of the corresponding classical theory\footnote{It is relevant to ask whether giving the quantum state and dynamics approximately fixes the classical solution that will be ``tracked'' in any valid classical situation, or whether different classical solutions could be possible within one classical situation.  For a semi-classical initial state one solution is approximately determined. In \cite{Kofler:2006,Kofler:2007} Kofler and Brukner argue that this need not always be the case; in fact, if one considers approximating the quantum mechanics with a probability distribution over classical states, the initial quantum state does not play a role in determining classicality at all.  Even in this case, some small set of initial measurements will approximately decide the classical history to be followed, so the distinction is not critical in the following and we can assume that one classical history is tracked.}, and if any subset of $M$ has the
same property\footnote{One might worry that, in theory, the last part of the condition is not obeyed by some macroscopic systems: without external interference some macroscopic systems can diverge from classical behaviour in surprisingly short times \cite{Habib:1998ai,Ballentine:2005,Schlosshauer:2006}.  Some
measurements in that case could make the system behave \textit{more} classically and it might seem that classicality need not be preserved if they are removed from the set $M$. However, this apparently does not lead to violation of the quantum/classical correspondence once a typical environment is added to the system, introducing decoherence effects.  At least in this limited sense, decoherence effects play the role of the missing measurements.}.
\end{definition}

This includes measurements in all bases.  A projection measurement $P_i \in M$ gives a range of possible values $\Delta_i$ for some observable $O_i$ at some time $t_i$.  A classical history $\gamma^{\text{cl}}$ will assign one value to the observable at that time, $O^{\text{cl}}_i(\gamma^{\text{cl}};t_i)$.  Not contradicting the classical results means
\begin{equation}
\exists \, \gamma^{\text{cl}} \; \text{s.t.} \; O^{\text{cl}}_i(\gamma^{\text{cl}};t_i) \in \Delta_i \quad \forall \, P_i \in M.
\end{equation}

The most obvious property of such a situation is that the measurements included will not be of arbitrary accuracy.  For instance, if $M$ is classical then, whatever the initial state, no two sharp measurements of anything can be in $M$ because the results would almost surely disagree with the exact values for any classical solution.

Any measurement, no matter how unsharp, does affect the state to some degree.  However the choice of whether to carry out a classical measurement will not affect the results of any of the others.  This follows from the above definition.  Consider a classical situation as above, with measurements $M$, and let $M'=M \backslash A$ be a smaller set of measurements.  By definition \ref{st:m},  all of the measurements in $M$ must confirm the classical predictions (with high probability), and the same is true of $M'$ alone, as it is a subset of $M$.  Because the measurements are two-valued, and only one outcome can confirm the classical prediction, this can only be true if the results of the measurements in $M'$ are the same, whether or not the other measurements in $M$ are performed.

Conversely, we can start with the following definition:

\begin{definition} \label{st:m2}
A set of measurements $M$ is classical (with respect to a particular
dynamics and state) if, with high probability, none of the
measurements contradict the predictions of some solution of the
corresponding classical theory , and the choice of whether to carry out any subset of the measurements $M$ does not affect the results of the other measurements.
\end{definition}

From this we can easily infer that all subsets of $M$ are also valid as the set of measurements in a classical situation (given the same initial state and dynamics). This gives an equivalent formulation to definition \ref{st:m}.

 This kind of classical situation is realised, for example, whenever the positions and angular momentum of solid macroscopic bodies like chairs and tables are approximately measured; looking forward to a theory of quantum gravity, any set of measurements so far performed in confirmation of General Relativity (along with a suitable state and dynamics) would constitute a classical situation.  It is interesting to ask when and how classical situations arise from more general quantum mechanical regimes, what degree of approximation is necessary for a given system, and so on.  Such studies can be found in some standard textbooks \cite{Peres:1995,Ballentine:1998} and this is still an active area of research (see \textit{e.g.} \cite{Kofler:2006,Kofler:2007,Habib:1998ai}).  In particular \cite{Kofler:2007} gives a specific and precise example in which a classical situation can be realised and the above claims are valid.

It is now straightforward to justify claim \ref{c:basis}.  It is necessary for the correspondence principle that classical situations exist in which the measurements in $M$ give enough information to approximately reconstruct a classical history, as we often do in reality.  Needless to say, given a classical history, \textit{i.e.} the configuration variables at all times, the value of other observables such as momentum and energy can be calculated.  The same is approximately true if the configuration as a function of time is only approximately known\footnote{The approximation necessary for the values of these other variables is a function of the approximation for measurements of distance.  We should not expect it to always be possible for the approximations to be small in terms of the average value, \textit{e.g.} for contrived variables such as exponentials of the energy \cite{Ashtekar:2005dm}.  This is acceptable physically, as it possible to satisfy the correspondence principle without measuring such contrived variables more accurately.}.  In classical situations, the results of reasonably unsharp measurements of, \textit{e.g.}, momentum could be approximated by a series of (approximate) position measurements.  It is sometimes said that ``classical measurements approximately commute'' in this context. We can conclude that all of the measurement outcomes that we find in a classical situation can be derived from a series of position measurements.  In those situations, there is no loss of generality incurred by talking only in terms of measurements in this one basis.  Similar reasoning would apply to more general systems than single particles.   In other words, all the measurements in our set $M$ can be re-expressed as a series of measurements one basis, which we can identify with the histories basis mentioned above.  This leads to claim \ref{c:basis}.

Having used it to justify the claims, we should note that definition \ref{st:m} above is not as strong as it might appear.  According to this requirement, it is possible for two sets of measurements $A$ and $B$ both to be good classical sets of measurements with respect to some state and dynamics, but for $A \cup B$ not to be. This is not a fault in the formalism.  Every measurement affects the state, and it is possible that performing one extra measurement could tip the balance and prevent another measurement from verifying the classical prediction, within our standards of high probability.

\subsection{The histories viewpoint}
\label{s:view}

For completeness, a stronger histories viewpoint should also be mentioned, although this has less relevance to the following discussion.  We can consider the properties of histories as giving a full account of all experimental outcomes, rather than just those in classical situations, albeit in a less direct fashion.  In his original paper \cite{Feynman:1948ur}  Feynman makes this point:

\begin{quote} The [SOH measurement] postulate is limited to
defining the results of position measurements.  It does not say
what must be done to define the result of a momentum measurement,
for example.  This is not a real limitation however, because in
principle the measurement of momentum of one particle can be
performed in terms of position measurements of other particles,
e.g. meter indicators.  Thus, an analysis of such an experiment
will determine what it is about the first particle which
determines its momentum. \end{quote}

With this in mind, it is possible to hold that measurements in other
bases are an unnecessary ``fifth wheel'' in quantum mechanics, which
need never be considered when corresponding our theory to reality.  In this view, concepts like momentum measurements are viewed as deriving from series of position measurements only in a classical approximation, and have no more fundamental meaning. One gives up Hilbert space basis symmetry as being of fundamental significance, perhaps in the hope of incorporating relativity more naturally into quantum mechanics. That is an expression of the histories view of quantum mechanics. It is even
possible to go further, and consider interpretations in which one
history (from the set corresponding to all measurement outcomes) represents reality (see \cite{Sorkin:1995nj,Kent:1997bc} and references therein). Such a history would indeed have all the properties needed to explain observations (although these approaches would arguably have to face the same challenges as De Broglie-Bohm pilot wave theories: does this notion of ``the real history'' serve all or any of the same purposes as its classical analogue, or is it itself a fifth wheel?).  In developing a feel for the histories approach, it may help to know that such views are possible; however, the comments below will not make direct use of this stronger interpretation.

Having reviewed the basic claims, we can now turn to some more specific questions.  The next two sections explain how symmetries and effective descriptions affect these claims, and thus show how claim \ref{c:approx} can be more than a trivial requirement.

\section{Symmetries and classical measurements}
\label{s:sym}

As has been emphasised, it is necessary for the correspondence principle that there exist classical situations in our quantum theory, allowing us to approximately reconstruct parts of a classical history, and thus find consistency with our experience of the classical world.  The classical theory may have symmetries, which we can verify by measurements in its domain of validity.  Does this affect the level of approximation that we can demand for claim \ref{c:approx}?

Let us consider the case of a macroscopic, rigid rotating body.  In everyday life, we are used to the fact that we can approximately measure all of the orientation and angular momentum components of such an object.  In other words, in a certain regime, we never find that measurement of the $z$-component of angular momentum of, say, a baseball, alters or precludes the measurement of the $y$-component.  Therefore it is tempting to make the following claim:

\begin{definition}
\label{d:generic}
Consider measurements $P$ and $P'=U^{-1}PU$, where $U$ is a unitary operator representing a symmetry transformation.  In a classical situation $\{|\Psi\rangle,U(t,t'),M\}$ that is ``generic'', if measurement $P$ is in $M$, then $\{|\Psi\rangle,U(t,t'),M'\}$ is also a classical situation, where $M'=M \cup P'$.
\end{definition}

However, it is not impossible to imagine a classical situation, as defined previously, in which this was not the case.  As noted at the end of section \ref{s:classmeas}, it is possible that for some state and dynamics, the sets of measurements $A$ and $B$ can both be classical, while the combined set $A \cup B$ is not.  For example, $A$ could be a measurement of the $z$-component of angular momentum with some high accuracy so that the situation is ``only just'' classical (\textit{i.e.} the result matches the classical one within our standards of high probability, but the probability is close to breaking those standards).  Then $B$ could be a rotated version of $A$, and carrying out both $A$ and $B$ could result in a probability to match the classical path that is outside our standards of high probability.

This situation is possible, but, as noted, not typical -- hence the insertion of the word ``generic'' in definition \ref{d:generic}.  To engineer such measurements on a baseball, we would have to measure the angular momentum components with tremendous accuracy, and perform the two measurements in such quick succession that environmental disturbance could not return the ball to a typical semi-classical state in the meantime.  It would be interesting to be more exact about this, but the main point stands without further analysis:  everybody is aware that we can, and almost always do, observe classical situations in which such difficulties do not arise, and verify symmetries of the classical theory in them.

In the light of this, it is reasonable to add definition \ref{d:generic}, as a strengthening of our definition of classical situations.  Again, we leave aside the question of how and when exactly such generic classical situations arise in standard theories.

Adding this claim is important, as it preserves the meaning of the symmetries in classical situations.  Take a classical theory and a symmetric state in that theory (the rotational symmetry of a region in a field, for example, or the global Lorentz symmetry of Minkowski space realised as a solution of GR).  In the domain of validity of the classical theory, we should be able to test the symmetry by performing a measurement, followed by several transformed versions of the measurement.  If the results of all measurements do not match, the symmetry is ruled out.  We can embed this classical case as a (generic) classical situation in a quantum theory, with the help of definition \ref{d:generic}.  We start with a symmetric quantum state and we can perform all measurements $P'=U_1^{-1}PU_1$, $P''=U_2^{-1}PU_2$, and so on.  All of these can be included in a classical situation, if one of them can, according to definition \ref{d:generic} (as long as the classical situation remains generic after the inclusion of each measurement).  Again, if results fail to match we can conclude that the symmetry is broken.  This is the situation in standard quantum mechanical theories.  For example, in a semi-classical state, the rotational symmetry of a region in a field could be verified in this way.

Without definition \ref{d:generic}, the situation is unclear.  Now, in the classical situation, it is possible to start with a symmetric quantum state, but obtain different results for the measurements $P$, $P'$ \textit{etc.}  There is nothing wrong with the symmetry of the quantum state; it is classicality that has failed instead in this case.  But what becomes of the classical test for symmetry?  Following the classical rules we would say that the symmetry is broken, but it is still possible that the initial quantum state was symmetric.  Indeed, performing a symmetry transformation on \textit{all} of the measurements, and repeating the experiment with an initial symmetric state, would of course produce the same results (statistically).

Further, consider the case in which we have no control over the first measurement $P$ -- we could even substitute for it an environmental decoherence effect.  In this case the failure of classicality caused by measurement would depend on the symmetry group element defined by $P$, something we could not even alter in different runs of the experiment.  The breaking of the symmetry by the uncontrollable disturbance of $P$, \textit{even though it did not affect the classical state being tracked}, could destroy the classical symmetry.  Emphasising the classical or quantum side, we could say: (\textit{i}) the notion of symmetry that we test classically is broken, and thus not properly captured by the symmetry of the quantum state, and (\textit{ii}) we cannot not trust our usual classical way of verifying symmetry.  In other words, claiming the existence of the symmetry would not be enough to predict that the usual classical tests would verify the symmetry.  Fortunately in all standard theories we can apply definition \ref{d:generic} and so this difficulty has never arisen, so far.  If it did arise, the onus would be on the inventors of the new theory to clarify matters.

The existence of these generic classical situations is also consistent with ``quasi-classical'' or ``disturbance uncertainty'' arguments.  The requirement is the following: the quantum uncertainty in some measurement of an object system $S$ should accord with the uncertainty produced by coupling a measurement system $T$ to $S$, if $S$ is treated as classical and $T$ is granted the uncertainty relations between conjugate variables dictated by quantum mechanics.  The Bohr-Rosenfeld argument \cite{Bohr:1933} is a venerable example of this type.  Failure of the classical symmetries by repeated measurement seems to be inconsistent with this, and might re-open some difficult conceptual problems for quantum mechanics.

In conclusion, such generic classical situations must exist in order for us to test (unambiguously) symmetries without leaving the domain of validity of the classical theory, \textit{i.e.} to test the symmetries using classical methods.  In the statement of the correspondence principle (section \ref{s:compare}), the classical situation referred to must be a generic one.  As in claim \ref{c:approx}, this should allow us to approximately reconstruct a classical history, symmetries and all.  So we can now demand something more from the approximation $\sim$:

\begin{claim}
\label{c:sym}
The level of approximation $\sim$ allowed in claim \ref{c:approx} should be consistent with the symmetries of the classical history being tracked.
\end{claim}

\textit{I.e.}, the approximation allowed for observable $O$ should be the same as for $U^{-1}OU$, where $U$ is a symmetry transformation operator.  In yet other words, $\sim$ should respect the symmetries of the classical theory.  This ensures that classicality can be preserved under symmetry transformed measurements.

\section{Effective descriptions}
\label{s:effective}

The variables that give the most useful description of a system at a macroscopic scale are often not those that we need to employ at a more fundamental level.  What is required is an effective (otherwise called coarse-grained or emergent) description of our fundamental theory.  This is the case, for example, whenever a hydrodynamical description of many particles is used.  Often, our classical theory is expressed in terms of effective properties, rather than the fundamental ones.  We may then want to discuss the quantum/classical correspondence principle in this effective language.  In quantum gravity, it has been hypothesised that the continua of GR may be effective descriptions of some discrete structure (see part \ref{p:qg}).  We will now discuss how this idea can be consistent with the claims made above.

Our starting point is a quantum theory as in the discussion of section \ref{s:hist}.  This is the fundamental theory, and should be complete in itself: in principle, it should supply all of our predictions, and all of the claims of the previous sections apply to it.  Certain measurements correspond to subsets of the history space $\Om$ as in claims \ref{c:outcomes} and \ref{c:main}.  However, the fundamental description of these subsets might be too complicated to use directly; we would like to relabel them in a way that is easily related to measurement outcomes.  The effective description is an efficient way of achieving this: we can express measurement outcomes in terms of subsets of some different ``effective'' history space $\Om'$, and then relate these to subsets of the fundamental history space.  That is, we can use some relation that, given an effective property corresponding to $\Gamma' \subset \Om'$ (\textit{e.g.} some value of a hydrodynamical mass density integrated over a region), outputs a fundamental property that is its expression in the fundamental theory (\textit{e.g.} some condition on configurations of many particles).  For convenience, the space $\Om'$ should contain all of the effective histories that we need to have a good effective description (\textit{i.e.} so that, when the effective description is valid, all of the measurement outcomes correspond well to effective history properties), and nothing else.

Let us formalise the idea of effective properties.  We will use a binary relation $\ap$ to associate fundamental histories $\gamma \in \Om$ to effective histories $\gamma' \in \Om'$ \footnote{In other words, there is a subset $E_\ap$ of  the Cartesian product $\Om \times \Om'$, and $x \ap y$ means $x \times y \in E_\ap$.}.  That is, $\gamma \ap \gamma'$ if $\gamma'$ is an effective description of the history $\gamma$.  Note that $\ap$ need not be a function.  There may be many effective descriptions of one fundmental history, and some fundamental histories may have no effective description at all (\textit{e.g.} a hydrodynamical description of a system may not cover all possible states of that system).
%NB: Got rid of $\Xi terminology, see previous drafts.

As required, every subset of the effective history space is related to a subset of the fundamental history space via $\ap$.  Consider an effective property $P$ and the associated set of histories $\Gamma'_P \subset \Om'$.  Let us say that a fundamental history $\gamma$ ``has effective property $P$'' if and only if \textit{all} effective descriptions of $\gamma$ have the property.  Thus, the subset of the fundamental history space associated to $P$ is
\begin{equation}
\Gamma_P=\{\gamma \in \Om : \gamma \ap \gamma' \Rightarrow \gamma' \in \Gamma'_P \}.
\end{equation}
This is the physically relevant concept:  if a measurement outcome definitely corresponds to $\Gamma'_P$, this implies that the fundamental history is in set $\Gamma_P$.

Using $\ap$, we can in principle calculate probabilities of measurement outcomes from the fundamental theory as before.  In practice approximations are used, and direct reference to the fundamental theory becomes unnecessary.  This effective description may be indispensable in practice, and can lead to a theory that looks, even qualitatively, quite different from the fundamental one\footnote{Note that similar statements could be made in the case of classical statistical mechanics, where is is obvious that all measurement outcomes correspond to properties of the fundamental histories.}.  However, the fact that it is in principle possible to relate histories in $\Om$ and $\Om'$ is the lynch-pin that holds the two descriptions together; without it the idea of an effective description could not be meaningful.  If a histories formulation is possible for the fundamental theory, then this way of describing the effective theory must also be possible, whatever other formulations might also be possible.

This represents no fundamental change in perspective from that of the previous section.  Using $\ap$, any question about effective properties corresponds to a question about fundamental histories, and the correspondence principle can be dealt with as before.  Conversely, there are no properties of the effective histories which cannot be re-expressed as properties of fundamental histories.  Again, there are no ``strongly emergent'' observable properties with no counterpart at the fundamental level.  The only difference is the extra layer of approximation introduced.

This conclusion follows from the claims of the previous section, and is not at variance with the perspectives coming from condensed state matter systems or quantum field theory.  It seems that the only field in which explicitly contradictory views are sometimes expressed is quantum gravity.  It is interesting to note that Wilson and Kadanoff's coarse-grainings, an essential ingredient of their renormalisation group techniques, are consistent with this picture \cite{Wilson:1974mb}.  There, a map is made between individual fine-grained lattice field configurations and individual field configurations on a coarser lattice, which gives the relation $\ap$ in the above terminology.  Roughly speaking, the quantum amplitude attributed to a coarse-grained configuration is calculated by summing over the fine-grained configurations that correspond to it.  A candidate $\ap$ is chosen on the basis that all fine-grained configurations that map to the same coarse-graining must be physically indistinguishable when probing on large scales.  This is an example of a situation in which a histories view has helped in the derivation and understanding of new physical concepts.

\subsection{Physical conditions on effective descriptions}
\label{s:cond}

There are some necessary conditions on $\ap$, if a good effective description of our theories in terms of $\Om'$ is to be recovered.  We have already said that we require \textit{all} members of $\Om'$ to be possible histories, so $\ap$ must be surjective (or ``right-total''): for all $\gamma' \in \Om'$, there must exist a $\gamma \in \Om$ such that $\gamma \ap \gamma'$.

Still, a fundamental history may contain \textit{less} information than an effective one: consider approximating a discrete set of atoms with a continuous medium, for example.  This can be accommodated by allowing $\ap$ to be one-to-many\footnote{In other terms, $\ap$ is not functional. A binary relation $R$ between domain $X$ and codomain $Y$ is functional (or right definite) if for all $x \in X$, and all $y, z \in Y$ it holds that if $xRy$ and $xRz$ then $y = z$. It is left-total if there exists a $y \in Y$ such that $x R y$ for all $x \in X$.  We have already noted that $\ap$ need not be left-total.}.  This is physically acceptable: because experiments cannot be carried out to arbitrary accuracy, not all histories in $\Om'$ need be physically distinguishable.  A fundamental history $\gamma$ can have more than one approximating effective history, as long as these effective histories are physically indistinguishable from one another.  We will write $\gamma'_1 \sim \gamma'_2$ if these two effective histories are physically indistinguishable.  Using this, the condition is:
\begin{equation}
\label{e:econd}
\forall\, \gamma \in \Om, \, \forall \, \gamma'_1,\gamma'_2 \in \Om', \:
\gamma \ap \gamma'_1 \, \text{and} \, \gamma \ap \gamma'_2 \,
\Rightarrow \,  \gamma'_1 \sim \gamma'_2.
\end{equation}
Here we have deliberately identified ``physical indistinguishability'' with the level of approximation allowed between classical and quantum histories, $\sim$.  This represents the minimal requirement that the accuracy of our measurements (\textit{i.e.} ability to distinguish the histories) must be at least as good as that used in the domain of validity of the classical theory.  If this condition were not true, we would be in the untenable situation of being able to measure for a property of a system, while denying that it has any meaning in our fundamental theory of that system; hence we would be unable to make any predictions about it from that theory.  These conditions can be combined with claim \ref{c:sym} of section \ref{s:sym}, which put a physical condition on the approximation relation $\sim$.

The requirements that we arrive at are the following:

\paragraph{Physical conditions on effective descriptions:} \textit{$\ap$ is surjective; the condition in eqn.(\ref{e:econd}) must hold;  in that condition, the relation of physical indistinguishability $\sim$ must respect any classical symmetries that we wish to preserve.}

\paragraph{}It is important to note that the symmetries mentioned here operate on the \textit{effective} histories $\gamma' \in \Om'$ that are the arguments of $\sim$ ; they may be absent or even meaningless on the fundamental level.  The symmetries need only have meaning at the effective level to connect with the classical theory.  For example, when gases are commonly referred to as translationally and rotationally invariant, this is a statement referring to the effective continuum level.  It has no counterpart at the fundamental level.

At this point the relevance to quantum gravity can be emphasised.  Say we have a classical theory to quantise.  Claim \ref{c:approx} gives a necessary condition on the history space, in order to maintain the correspondence principle.  This seems like a weak condition, and in standard cases our quantisation procedures ensure that the condition is satisfied.  However, say we believe that the best route is to propose some more fundamental history space which is only linked to the classical histories via an effective description $\ap$.  In this case claim \ref{c:approx} leads to the above conditions, and may become non-trivial.  Further investigation is needed to ensure the correspondence principle can hold.  This can be taken care of once something is known about $\Omega$ and $\ap$; we need not wait for the full definition of our quantum theory.

\part{Consequences for quantum gravity}
\label{p:qg}

How would gravity fit into the histories framework?  The review in section \ref{s:hist} started with the example of non-relativistic particles; histories as described there were functions of some fixed time co-ordinate.  In GR such a time co-ordinate does not have a physical meaning independent of some co-ordinatisation.  However, in a more general view, the quantum histories need only be some unsmooth generalisation of the classical histories, and the discussion above would carry over unaltered.  For example, theories of the time-reparameterisation invariant particle can be made in which the histories are not functions of time, and a theory of gravity might work in a similar way \cite{Hartle:1992as}.  It may even be possible to use the standard idea of histories as a function of time, given some gauge fixing.

Several prominent QG approaches have been built on the presupposition that a histories framework will remain valid here: Regge calculus approaches \cite{Williams:2006kp} including Causal Dynamical Triangulations \cite{Ambjorn:2006jf,Ambjorn:2005jj,Ambjorn:2005qt}, as well as the related spin-foam models \cite{Oriti:2001qu,Perez:2004hj} conjectured to include a spacetime formulation of loop quantum gravity \cite{Perez:2004hj}, and causal sets \cite{SpacetimeAsCS,Henson:2006kf,Dowker:2006wr} (See also \cite{Forks} for a discussion of the advantages of a histories approach in quantum gravity).  Any approach based on condensed matter or quantum-computational analogues, since they use more-or-less standard quantum theory, could be cast in a histories language.  Whether the histories framework for gravity is, as in standard theory, equivalent to a canonical formulation, or whether it would be more fruitful in this case to go beyond the canonical form altogether, is a debated issue.  The only assumption below is that a histories view is possible.  Indeed, if a generalised quantum theory was made which somehow did \textit{not} allow a histories formulation, this would itself raise the question whether too much of the standard quantum picture had been given up for the theory to be physically reasonable.  The whole physical interpretation of such a theory, especially the correspondence principle, would have to be thoroughly re-evaluated before it could be accepted.

\section{Applying the histories approach}
\label{s:qg}

In claims \ref{c:main} and \ref{c:approx}, we have seen that there are physical constraints on the quantum history space, coming from the requirement that our quantum theory satisfies the correspondence principle.  For a quantum gravity theory to have the correct classical approximation, there must be some histories in $\Om$ which have all the properties that we expect to be able to observe in a classical situation.  A classical situation here (real or hypothetical) would be one in which General Relativity is verified by experiment and observation.  It could, for instance, include \text{all} measurements that have actually been made that are relevant to GR.  The following is an adaptation of claim \ref{c:approx} to quantum gravity.
\begin{claim}
\label{c:qg}
Consider a quantum gravity theory that is consistent with the histories framework.  To be physically reasonable, some histories of the theory must approximate to solutions of general relativity\footnote{The argument leading to claim \ref{c:qg} requires that the histories be written in some reasonable histories basis, corresponding to some reasonable basic observable of the classical theory.  This is the case for spin-foams, where the histories are supposed to correspond to triangulations of spacetimes with labelled faces.  These variables are supposed to relate to those of GR in a direct way.}.
\end{claim}
The appropriate level of approximation should obey the same conditions as for claim \ref{c:approx}.  Using effective descriptions as in section \ref{s:effective}, the fundamental histories do not have to be spacetimes (meaning diffeomorphism classes of Lorentzian manifolds) in order to satisfy claim \ref{c:qg}.  Also, from claim \ref{c:sym}, we might add that \textit{the level of approximation required in claim \ref{c:qg} for each observable should be consistent with the symmetries of the spacetime being tracked}.

Claim \ref{c:qg}, though derived from fairly uncontroversial grounds in standard quantum theory, may have interesting implications, particularly for two interrelated issues: ``emergent gravity'', and the fate of Lorentz symmetry in quantum gravity.

\section{Lorentz invariance}
\label{s:lorentz}

Local Lorentz Invariance, or more specifically the question of
whether this physical principle will be maintained or broken at
the Planck scale, has become a much debated topic both in the
theory and phenomenology of quantum gravity.  On the
phenomenological side, constraints on violation of Lorentz
invariance (termed ``Lorentz violation'' below) are becoming ever
more stringent \cite{Amelino-Camelia:2004hm,Jacobson:2005bg}, and strong arguments have been made that Planck scale Lorentz
violation is incompatible with current observations, without (at
best) additional fine-tuning being introduced
\cite{Collins:2004bp,Collins:2006bw}.  With this progress, it
becomes increasingly important for the theoretical side to provide
predictions, at least of a heuristic nature: in any particular
theoretical framework, what becomes of Lorentz invariance?  Is it
possible to maintain it, or must it be broken at some scale?

It is well-known that waves travelling on lattices violate Lorentz
invariance, and this is often used as an argument against
fundamental discreteness.  Some Loop Quantum Gravity (LQG) based
arguments have been made to support this expectation
\cite{Gambini:1998it,Sahlmann:2002qj,Sahlmann:2002qk}, and some
further discussion of the methods used is given in
\cite{Bojowald:2004bb}. However, there are conflicting arguments
for local Lorentz invariance in LQG \cite{Rovelli:2002vp}, and
also some \cite{Livine:2004xy, Freidel:2005me} for the ``third way'' of Doubly Special Relativity (DSR) \cite{Amelino-Camelia:2005ne} in which the Planck length or Planck energy is introduced as an invariant
scale along with the speed of light, deforming the Lorentz
transformations (an idea that is criticised in
\cite{Sudarsky:2005ua}). If the spin-foam quantum gravity program
is to provide a path-integral formulation of LQG, then it might be
expected that the argument which prevails in the LQG program will
also prevail for these models, and \textit{vice-versa}.

Here the question will be more closely examined.  Planck-scale lattices do violate Lorentz invariance, when considered as fixed backgrounds rather than in a quantum sum.  But does this have any physical significance for quantum gravity theories?

Regge triangulations are in a strong sense lattice-like.  They are co-ordinate free, meaning that the metrical information is all in the form of lengths between vertices, and the lattice connectivity (or in some generalisations, areas of triangles and/or angles).  Having said that, the classical theory of Regge calculus is not applied with finite lattice spacing, except as an approximation; the limit of small edge lengths must be taken to recover general relativity and all its continuum symmetries.  It has never been claimed that, with a finite lattice spacing, a space of Regge triangulations could approximate a reasonable space of spacetimes without violating local Lorentz symmetry (this will be discussed more precisely later).

Thus it would not be surprising if a quantum theory built on such Regge triangulations violated Lorentz symmetry, unless a continuum limit were taken.  After all, the classical theory being quantised is Lorentz violating.  However, in some approaches it is claimed that the problem can be overcome by a quantum sum over many triangulations.  The program to which this issue is of most relevance is spin-foams, where fundamental ``Planck-scale'' discreteness is sometimes claimed.  In other Regge calculus inspired approaches the lattice spacing is taken to zero.  In the causal set approach, the histories are not lattice-like in this sense.  They are claimed to be discrete \textit{and} to have local Lorentz symmetry, in any situation in which a continuum approximation is valid \cite{Dowker:2003hb,Bombelli:2006nm}.

\subsection{A condition for symmetry preservation in discrete quantum gravity}
\label{s:lorentzqg}

In theories like spin-foams, the spacetimes of GR (perhaps with additional fields) are effective descriptions of the fundamental discrete structures.  When assessing such theories from the histories perspective, we should therefore apply the discussion of section \ref{s:effective}, and the conditions on effective descriptions developed in section \ref{s:cond}.  We have the fundamental histories $\Om$ and the effective histories $\Om'$.  There should be a binary relation $\ap$ such that $\gamma \ap \cM$ if the spacetime $\cM$ is an effective description of the history $\gamma$.  To apply the conditions from section \ref{s:cond} to the case of quantum gravity:

\paragraph{Physical conditions on emergent spacetime:} \textit{$\ap$ is surjective;
\begin{equation}
\label{e:haupt}
\forall\, \gamma \in \Om  \text{ and } \cM_1,\cM_2 \in \Om', \quad
\gamma \ap \cM_1 \, \text{and} \, \gamma \ap \cM_2 \,
\Rightarrow \,  \cM_1 \sim \cM_2;
\end{equation}
here, the relation of physical indistinguishability $\sim$ must respect all symmetries of GR that we wish to preserve.}

\paragraph{} To recap, the last clause follows if we accept the discussion of section \ref{s:sym} on symmetries.  As before, the symmetries are not applied directly to the discrete structures themselves.  These symmetries, such as local Lorentz symmetry, have their first definition at the level of the approximating continua.  That is where they are properly applied and tested.

This is an important and non-trivial condition, which would be difficult to formulate without the histories perspective given in earlier sections.  It depends only on $\Om$ and $\ap$ for each theory (suitable definitions of $\Om'$ and $\sim$ should be fixed by GR), and so the condition is in some sense only kinematical; we do not need to solve the theory or even know the full dynamics to test it.

In causal set theory this condition is known as the \textit{hauptvermutung} or ``fundamental conjecture'' and justifying it is taken as a crucial task.  In that case, a discrete/continuum correspondence can be defined that does indeed respect the symmetries of GR in the effective continuum description (\textit{i.e.} it involves no special choice of co-ordinates, frames, foliation, \textit{etc.}).  As well as this, there is now compelling evidence for the \textit{hauptvermutung}: timelike distances \cite{Brightwell:1990ha}, dimension \cite{Meyer:1988,Myrheim:1978ce,Reid:2002sj} and topology \cite{Major:2006hv} can all be approximately reconstructed from a causal set.

In the spin-foam approach, no attempt has been made to justify a condition of the form of eqn.(\ref{e:haupt}).  There is no explicit proposal for $\ap$.  There is no discussion in the spin-foam literature explaining why the condition is safely to be ignored, leaving us to reconstruct the probable arguments.  It may be due to disagreements on some statements made that led up to (\ref{e:haupt}), for example in section \ref{s:cond}.  Disagreement there could lead one to refute that the condition is necessary in order to produce GR in a semi-classical limit.  On the other hand, it may be due to some implicit confidence that such a condition could be satisfied by any reasonable discretisation of spacetime, with a good choice of $\ap$.

In contrast, I conjecture that the condition must fail in spin-foam theories, for any choice of $\ap$ of the expected form, if fundamental Plank-scale discreetness is to be retained.  By ``expected'' I mean that vertices will in some sense correspond to spacetime points, and so on for higher dimensional simplices, as suggested by the Regge calculus on which the spin-foams approach relies.  The conjecture will not be proved in this paper.  However, some reasons will be given to doubt that the physical conditions given above can be satisfied for spin-foams.  After this, the symmetry arguments from section \ref{s:cond} will be briefly reviewed in the QG context to answer the other possible disagreement.

\subsection{The problem with lattices}
\label{s:sfargument}

After the above discussion, we can reassess the question: is it significant that lattice-like histories violate Lorentz symmetry, when taken individually?  First we should say what this means.  Let us concentrate on Minkowski space, in which Lorentz symmetry is global, bearing in mind that this is a good approximation to the geometry of significant regions of the actual universe.

Let us first consider a fixed background case.  The lattice can be considered as an abstract graph up to isomorphisms, perhaps with length-labelled edges.  Consider  a square (or diamond) lattice $L$ with vertices $v_L$, and fields $\phi_L: v_L \longrightarrow \Reals$ on the lattice, and let each field configuration be a member of  $\Om$.   Now consider approximating these by real scalar fields on Minkowski space $\phi_\Mink \in \Om'$.  This kind of discretisation is very common, and the definition of $\ap$ is simple.  It can be written in a way that explicitly invokes a preferred co-ordinate frame in Minkowski space, but this is not necessary.  Another form that does not reference a special frame can be written.  The rough form of such an $\ap$ is given in the following:

\paragraph{} $\phi_L \ap \phi_\Mink$ iff there exists some embedding $\iota : v_L \rightarrow \Mink$ (i.e. a map from vertices of the graph $L$ to points of $\Mink$) such that $\forall x \in v_L$, $\phi_L(x) \approx \phi_\Mink(\iota(x))$, and such that the lattice distances ``match'' those in $\Mink$.

\paragraph{} There is no need to be specific about the type of matching of distances to be used, as the following argument is not sensitive to these details.  It could be exact or approximate.  We want to discreteness scale to be ``Planckian'' in some sense, in accord with familiar arguments in QG; again we need not be specific.  We could say, for example, that the embedding should include some Plankian order of vertices in a region of unit volume.

Such a discrete-continuum correspondence $\ap$ must be used when the continuum limit is not taken, but it does not satisfy our physical conditions on emergent spacetime.  Although there is no use of a frame selected \textit{ab initio},  the embedding introduces a special lattice frame, and so the correspondence obviously depends on a frame.  Some properties of the continuum histories are missing from the lattice histories.

Consider a field configuration $\phi_{\Mink \, 1}$ in which that there is one plane wave travelling.  The lattice can only support a limited bandwidth and so the above definition for  $\ap$ cannot distinguish between such a field configuration and another, $\phi_{\Mink \, 2}$  in which there is a second wave with frequency vastly differing from the first.  This is straightforward to see:  in whatever embedding was used to show that $\phi_{\Mink \, 1} \ap \phi_L$, we add a wave with the lattice spacing as its wavelength to form $\phi_{\Mink \, 2}$, and thus find that the same embedding can be used to show  $\phi_{\Mink \, 2} \ap \phi_L$.  But, in contradiction of our physical condition on emergent spacetime given in equation (\ref{e:haupt}), it cannot be that $\phi_{\Mink \, 1} \sim \phi_{\Mink \, 2}$.  Whatever the details of our approximation $\sim$ are, the presence or absence of a wave of arbitrary amplitude -- whatever the frequency happens to be in any particular frame -- is surely something that should be considered physically distinguishable, \textit{if} $\sim$ is to be Lorentz invariant.  It is an uncontroversial claim that such a lattice kinematics inevitably leads to Lorentz symmetry breaking; it can only work with a Lorentz violating condition for $\sim$.  (It is interesting to go over this argument in the case of Euclidean symmetry; in this case we find the above correspondence to be quite adequate).

In this simple case, symmetry can be restored by passing to a random lattice.  In the background dependent case, if points are distributed according to a Poisson process this eliminates the lattice frame \cite{Bombelli:2006nm}.  But the positions of the points cannot be allowed \textit{the slightest} uncertainty. For, if we were to randomly move them in a ball-shaped neighbourhood, then the frame defined by these balls would re-introduce Lorentz breaking.  And the Lorentzian analog of a neighbourhood of $x$, \textit{i.e.} those points with Lorentzian distance $\epsilon$ of $x$, is of infinite volume for any non-zero $\epsilon$ in Minkowski space.  In a theory in which lengths are dynamical, therefore, it is difficult to imagine this random lattice trick working (unless one uses an infinite valency ``lattice'' like a causal set).

It will be left to another place to fully explain how the approximation fails, and how similar the background independent case is.  However, it is not hard to see that similar problems arise in that case.  The above correspondence can be easily extended to fields over many lattices, and many spacetimes, instead of just one; replace $\Mink$ with $\cM$ in the above condition.

None of the ``rules'' of fundamental discreteness were tacitly broken here.  The discrete structure is a ``co-ordinate free'', abstract graph and requires no background manifold to live on.  No background geometry is presupposed.  However, a continuum spacetime must come into the discussion when we consider continuum approximations, and that is where the notion of embedding enters.  Even then, the requirement is only that \textit{some} embedding exists, which is a (possibly edge labelled) graph invariant and co-ordinate invariant requirement.  Hence this conclusion cannot be evaded on the grounds that the lattice or the discrete/continuum approximation is not ``background independent'' enough.  Given the need for a discrete/continuum correspondence $\ap$ to exist, no \textit{more} background independent realisation of the idea is possible.

These considerations, at least, cast doubt on the idea that spin-foams or any lattice-like structure based on finite-valency graphs can satisfy the condition of equation (\ref{e:haupt}).  But we should also re-examine the question whether such a condition is necessary in this context.

\subsection{Examining the condition}
\label{s:excond}

If the above conjecture is correct, a spin-foam may have some properties corresponding to those of Minkowski space, but will lack some others that it would need to be Lorentz invariant.  According to the claims, in a spin-foam theory it would then be impossible to measure for all of these properties in a classical situation.  Despite this apparent problem, arguments have been made in favour of the recovery of local Lorentz invariance in spin-foam models.  These are briefly given in this section.

In \cite{Rovelli:2002vp}, the point is made that, in a semiclassical state of Loop Quantum Gravity representing Minkowski space, it is possible for two sharp measurements related by a boost to be non-commutative, just as measurements of components of angular momentum are non-commutative in standard quantum theory.  In the histories language, this means that both cannot exactly correspond to properties of a history.  These sharp measurements would not be classical, and so they are not directly relevant to the above problem.  However, it has been argued\footnote{\label{f:carlo} The argument is based on my understanding of some comments made by Carlo Rovelli and Daniele Oriti in private communication.} that some caveats in the discussion of classical measurements can be exploited to evade the problem raised in the previous section.

The argument is that, in quantum gravity, the strong idea of the ``generic classical situation'' put forward in section \ref{s:sym} need not apply.  This undermines claim \ref{c:sym}, and in turn, one of the physical requirements on effective descriptions in the argument above.  Without generic classical situations, there is no reason why the discrete/continuum correspondence should respect the symmetries of GR.  Properties of the histories that are lacking are simply not behaving classically.

Consider two coarse-grained measurements $P$ and $P'$ related by a boost.  In a state supposed to represent Minkowski space, a measurement $P$ could be measured in one Lorentz frame and considered part of a classical situation -- but could carrying out both $P$ and $P'$ yet be disallowed in a classical situation?  Well, this would be consistent as long as these boost-related properties are never both present in a realistic set of classical measurements.  This is a view of Lorentz symmetry sometimes put forward for spin-foam quantum gravity.  Thus is it claimed that a semi-classical, Lorentz invariant vacuum state for quantum gravity could still exist in such a theory, and give rise to physically acceptable classical situations.  The claim is the following: it is not the symmetry which fails in this case, but classicality.

Let us imagine the earth's frame to be the one near which most measurements are carried out.  In this scenario, for example, there may be no semi-classical state in the theory describing both the everyday events on and near the earth, and a photon of trans-Planckian frequency (with respect to the earth) moving on a classical background near the earth.  The only ``preferred frame'' here, it is argued, is the one in which measurements are carried out.  In a fuller treatment, that is merely a consequence of the configuration of the measurement apparatus.

There are a number of criticisms that could be made of this argument.  Many reasons have already been given for the requirement of generic classical situations in section \ref{s:sym}.  As we claimed there, the existence of such generic classical situations is not usually doubted in any standard quantum theory.  Any new theory that did not share this property would therefore be a radical departure; since one of the underpinnings of the classical approximation may have been compromised, this would necessitate a thorough evaluation of the plausibility of this new kind of quantum theory.  As discussed before, this change may also be at odds with quasi-classical reasoning about quantum theory.

But the main problem would, again, be the interpretation of experiments designed to test the symmetry.

When we talk of Lorentz violation we are primarily interested in making predictions for the many relevant observations being undertaken.  For example, a Lorentz invariant theory should return null results for photon time-of-flight experiments, (see \textit{e.g.} \cite{Ellis:2005wr}).  These observations are a test of a classical symmetry: they use classical methods as discussed in section \ref{s:sym}, and their interpretation does not directly invoke the quantum nature of spacetime. The connection between this phenomenological Lorentz invariance, and the Lorentz invariance of a quantum state, is destroyed if we do not demand generic classical situations, as has already been argued in the general case.

Say we were given this semi-classical state, and wanted to test its Lorentz invariance.  Consider some set of measurements designed to do this, $A$, and a set $A'$ in which all of those measurements have been Lorentz transformed (with the same transformation).  We could carry out measurements $A$, or $A'$, and we would of course find the same results (statistically) in either case.  But in reality, we cannot control all of the measurements made (a case discussed in section \ref{s:sym}).  No-one doubts that the geometry of the universe looks semi-classical near the cosmological frame: it must be decohered by environmental effects, and this semi-classicality is confirmed by literally every experiment.  Therefore, even if classically we are in Minkowski space (at least approximately and in some region), the quantum state has already lost the expected symmetry.  What, then, do we predict for the current ``Lorentz violation'' tests?  We could claim that far from the cosmic frame, it is not the symmetry but classicality which could fail, but that does not help us to predict anything.  Without generic classical situations, \textit{The Lorentz invariance of the state gives no reason to predict null results for these tests.}  Thus this type of ``Lorentz invariance'' without the generic classical situation does not connect with the phenomenological criterion of Lorentz invariance currently under scrutiny.  That should be the final criterion for Lorentz invariance.

Thus, claiming a Lorentz invariant quantum state without any further assumptions is not sufficient to rule out positive results for the real Lorentz violation tests.  However, the physical conditions on emergent spacetimes given in section \ref{s:lorentzqg} do give a necessary condition for the recovery of classical symmetries by all possible tests of this type.  If no spin-foam has the necessary properties to preserve Lorentz invariance in a classical situation, such a theory should be considered as giving rise to the type of Lorentz violating scenario currently being tested for.

More on the applying the histories approach to emergent geometry scenarios, including condensed matter analogs, is included in appendix \ref{s:emergentgeom}.

\section{Conclusion}

It is well-known that many classical intuitions fail in quantum mechanics.  Despite this, we have discussed one idea that can be preserved: a correspondence between measurement results and properties of histories.  From its beginning, this histories perspective has been especially useful in describing the emergence of classical behaviour in quantum systems, and this has been applied to quantum gravity.  The existence of many views on quantum mechanics is useful for progress in new theories, as well as for new insights on standard physics.  When a result is brought to light by application of one perspective, as above, other approaches can sometimes be used to further examine it.  However, the question arises in each such case:  does switching to a different perspective actually show a loop-hole to the result, or only obfuscate the result?

Consideration of Feynman's histories approach has revealed something about the possible phenomenology of quantum gravity theories.  Lorentz violation has been the most consistently pursued avenue for detecting QG effects, and it has been conjectured above that (fundamentally discrete) spin-foam theories cannot evade Lorentz violation, using the histories view.  In particular, it was argued that this conclusion could not be evaded by invoking a quantum sum over lattices.  To extend the work, more evidence could be given for the conjecture of section \ref{s:lorentzqg}, that spin-foams cannot satisfy the physical condition (\ref{e:haupt}) that is necessary to preserve Lorentz invariance.  However, as mentioned already, since there is no current evidence that spin-foams \textit{can} satisfy the condition, the onus is arguably on supporters of this view to provide evidence for it (or to undermine the condition (\ref{e:haupt}) in some other way).

It would also be of use to study the properties of individual histories in plausible spin-foam path integrals, and to give more details of a correspondence relation $\ap$ to continuum spacetimes.  This would help to decide with more precision which metric properties are not well approximated (in histories that approximate a well-behaved spacetime near one frame).  This would help to determine possible observational consequences.

The idea of deformed Lorentz symmetry \cite{Amelino-Camelia:2005ne} is also of possible relevance in this discussion.  It may be possible to square the above conclusions of undeformed Lorentz violation with this idea.  No way of doing this is immediately apparent, and so this presents another avenue for study.

As already commented, the consequences of \textit{the possibility} of a histories viewpoint cannot be ignored if one wishes to preserve standard quantum theory.  In the converse case, one would lose the standard arguments for the emergence of the classical approximation; a new justification for a correspondence principle would have to be arrived at.  This presents a great challenge for any ``trans-quantum'' theory.  But even so, perhaps such radical steps are exactly what is necessary to reach the next level in fundamental theory \cite{Hardy:2006uc,Doring:2007ib}.

\section*{Acknowledgements}

The author would like to thank Johannes Kofler for helpful comments on the classical limit of quantum mechanics, Harvey Brown for comments on gravity, and many participants at the Loops conference series for discussion of their views on Lorentz invariance, particularly Carlo Rovelli and Danielle Oriti.  This work also benefitted from conversations with Fay Dowker and Rafael Sorkin.

\bibliographystyle{h-physrev3}
\bibliography{refs}
\appendix

\section{Emergent Geometry}
\label{s:emergentgeom}

We have already discussed the idea of considering spacetime as an approximation to some other more fundamental entity, and its place in some popular approaches to quantum gravity.  This idea has been much influenced by condensed state matter analogs, in which relativistic field dynamics can appear as a hydrodynamical approximation to an underlying atomic dynamics \cite{Barcelo:2005fc,Visser:2007du,Volovik:2006ft}.  This appendix reviews and comments on some ideas that have been inspired in this way, using the histories perspective developed above.

How far can we go in removing geometry from the fundamental level of our theory?  Claim \ref{c:qg} clarifies what is possible and what is not.  Following the discussion of section \ref{s:effective}, it is indeed possible to have a history space $\Om$ for quantum gravity that is not merely a reasonable set of Lorentzian manifolds $\Om'$, or unsmooth generalisations thereof.  However, there must be some way to approximate at least some of these histories with continuum spacetimes.  We need a correspondence relation $\ap$ between the history spaces that satisfies certain conditions.  Examples include (\textit{a}) causal sets, in which a reasonably explicit discrete/continuum correspondence has been given \cite{Henson:2006kf}, and (\textit{b}) the ``quantum graphity'' model \cite{Konopka:2006hu,Konopka:2008hp}, which implicitly invokes a discrete/continuum correspondence when the occurrence of lattice-like graphs is said to be sign of emergent geometry (some form of graph-connectivity/distance correspondence is being presupposed).  Not all histories or configurations must have geometrical approximations, but some must.  This is as far from geometrical degrees of freedom as it is possible to get if the theory can be cast in a histories framework.

The effective degrees of freedom may have an unexpected relation to those at the underlying level.  For instance, if the fundamental structures are graphs, it need not be true that the graph distance between two vertices is directly related to the length between two corresponding points in the approximating manifold \cite{Markopoulou:2007ha}.  This would amount to a counter-intuitive property of $\ap$.  However, if we agree with the discussion of section \ref{s:effective}, we can conclude that all problems in relating fundamental and effective degrees of freedom can be expressed as problems with defining and interpreting $\ap$; there is no further ambiguity introduced when carrying out the quantum sum.  The properties of spacetime that we can measure (which are, of course, classical) are nothing but the common properties of some set of fundamental histories.

A histories perspective like this can help to explain some properties of known emergent geometry scenarios.  Consider the condensed matter ``analog models'', in which relativistic fields emerge as an effective description, \textit{e.g.} of density peturbations in a fluid \cite{Barcelo:2005fc,Visser:2007du,Volovik:2006ft}.  In these scenarios, there is a preferred frame in the fundamental description: the average rest frame of the fluid.  The atomic/continuum correspondence makes explicit use of this frame.  The effective, Lorentzian description is not perfect, and it breaks down when considering high frequency modes in the field (with respect to that underlying rest frame) \cite{Weinfurtner:2005jd}.  Why is this?  Using the histories perspective, there is an obvious conjecture:  the fundamental description, \textit{e.g.} in terms of molecular worldlines, does not have any properties corresponding to field modes of smaller wavelength than the molecular separation.  The effective description in terms of a mass density field cannot work at that scale.  The discrete/continuum correspondence $\ap$ picks out a time direction in the effective spacetime, and violates the physical conditions necessary for Lorentz symmetry to be maintained.  This seems a compelling conjecture, as we have already seen how difficult it is to design discrete/continuum correspondences $\ap$ that respect Lorentz invariance.
\begin{comment}
A possible loop-hole here is that the underlying description may not really be that discrete.  In a Bose-Einstein condensate there are discrete atoms but the best description might not be world-lines but continuous scalar fields.  In that case one might imagine finding a properties that correspond to all modes in the field for $\ap$.  In that case the conjecture is maybe too strong, so you need something more sophisticated but along the same lines, bringing in the dynamics.  Find out what happens to these high frequency modes -- they could even be totally inert, which would be something like the original conjecture.  For phonons on a lattice the conjecture seems almost immediately correct.
\end{comment}

Some hope to recover GR in a similar way: as an effective description of some system with a fundamental cutoff and a preferred time direction. % (this includes \cite{Markopoulou:2006qh,Markopoulou:2007qg} discussed below).
It has been argued that the symmetries, as measured by an observer who is internal to the system at the effective level, need not be affected by the fundamental preferred frame \cite{Dreyer:2006pp, Dreyer:2007ws}.  But in the case of Lorentz symmetry, there is no example of this known in condensed state matter analogs -- the effective symmetry always falls victim to the fundamental preferred frame at high energies.  The above conjecture would explain this, and show how generic the problem is.  The point is that this type of fundamental history space $\Om$ and fundamental/effective correspondence $\ap$ do not obey the physical conditions of sections \ref{s:cond} and \ref{s:lorentzqg}.  This would arguably be a stumbling block for any such ``emergent gravity'' theory,  with Planckian discreteness and a preferred fundamental time direction, if local Lorentz symmetry is not to be broken at high energies.

While it is true that effective symmetries need not be present or meaningful on the fundamental level, that is not to say that properties at the fundamental level are completely irrelevant to the recovery of effective symmetries.  We have already seen in detail how they relate, via $\ap$, in sections \ref{s:effective} and \ref{s:lorentzqg}.  In all of the known analog models, however $\ap$ would look in detail, it is clear that it links the preferred underlying time direction with a preferred time direction in the effective spacetime. According to the conjecture above, Lorentz violation could only be evaded if this troublesome property of $\ap$ could be removed; \textit{i.e} if the correspondence $\ap$ could be made fully Lorentz invariant.  In causal sets, inspired by a histories viewpoint, this property is ``designed in'' from the start.  This conclusion cannot be evaded by a preference for some other viewpoint on emergence; it is only made less obvious.

\end{document}